\documentclass[preprint]{aastex63}
\usepackage{amsmath}
\usepackage{multirow}

\newcommand       \Angstrom     {\,{\rm \AA}}
\newcommand       \AU           {\,{\rm AU}}

\newcommand       \g            {\,{\rm g}}
\newcommand       \K            {\,{\rm K}}

\newcommand       \s            {\,{\rm s}}

\newcommand       \yr           {\,{\rm yr}}

\newcommand       \um           {\,{\rm \mu m}}
\newcommand       \mum           {\,{\rm \mu m}}

\newcommand       \simali       {\sim\,}

\newcommand       \km           {\,{\rm km}}

\newcommand	      \rh           {r_{\rm h}}
\newcommand       \amin         {a_{\rm min}}
\newcommand       \amax         {a_{\rm max}}
\newcommand       \BV           {\left(B-V\right)}
\newcommand       \VR           {\left(V-R\right)}

\received{\today}
\revised{\today}
\accepted{\today}
\submitjournal{ApJ}

\shorttitle{Dust and Volatiles in Comet C/2019 Y4 (ATLAS)}
\shortauthors{Zhao et al.}
\graphicspath{{./}{figures/}}

\begin{document}

\title{Dust and Volatiles in the Disintegrating Comet C/2019 Y4 (ATLAS)}




\author[0000-0003-4936-4959]{Ruining Zhao}
\affiliation{CAS Key Laboratory of Optical Astronomy, National
  Astronomical Observatories, Chinese Academy of Sciences, Beijing
  100101, China; {\sf rnzhao@nao.cas.cn; jfliu@nao.cas.cn}}
  
\affiliation{School of Astronomy and Space Sciences, University of Chinese Academy of Sciences, Beijing 100049, China}

\author[0000-0002-1119-642X]{Aigen Li}
\affiliation{Department of Physics and Astronomy, University of
  Missouri, Columbia, MO 65211, USA;
  {\sf lia@missouri.edu}}

\author[0000-0002-5033-9593]{Bin Yang}
\affiliation{Instituto de Estudios Astrofísicos, Facultad de
  Ingeniería y Ciencias, Universidad Diego Portales, Santiago, Chile;
  {\sf bin.yang@mail.udp.cl}}

\author[0000-0003-3603-1901]{Liang Wang}
\affiliation{Nanjing Institute of Astronomical Optics \& Technology, Chinese Academy of Sciences, Nanjing 210042, China}
\affiliation{CAS Key Laboratory of Astronomical Optics \& Technology, Nanjing Institute of Astronomical Optics \& Technology, Chinese Academy of Sciences, Nanjing 210042, China}

\author[0000-0003-3271-9709]{Huijuan Wang} 
\affiliation{CAS Key Laboratory of Optical Astronomy, National Astronomical Observatories, Chinese Academy of Sciences, Beijing 100101, China}
\affiliation{School of Astronomy and Space Sciences, University of Chinese Academy of Sciences, Beijing 100049, China}

\author{Yu-Juan Liu}
\affiliation{CAS Key Laboratory of Optical Astronomy, National Astronomical Observatories, Chinese Academy of Sciences, Beijing 100101, China}

\author[0000-0002-2874-2706]{Jifeng Liu}
\affiliation{CAS Key Laboratory of Optical Astronomy, National
  Astronomical Observatories, Chinese Academy of Sciences, Beijing
  100101, China}
\affiliation{New Cornerstone Science Laboratory, National Astronomical Observatories, Chinese Academy of Sciences, Beijing 100012, China}
\affiliation{School of Astronomy and Space Sciences, University of Chinese Academy of Sciences, Beijing 100049, China}
\affiliation{Institute for Frontiers in Astronomy and Astrophysics, Beijing Normal University, Beijing, 102206, China}


\begin{abstract}
C/2019 Y4 (ATLAS) is an Oort cloud comet
with an orbital period of $\simali$5895$\yr$.
Starting in March 2020, its nucleus underwent disintegration.
In order to investigate the gas and dust properties
of C/2019 Y4 (ATLAS) during its disintegration,
we obtained long-slit spectra at 3600--8700$\Angstrom$
and $BVRI$ multi-band images
with the Xinglong 2.16-Meter Telescope in April 2020.
Our observations revealed that C/2019 Y4 (ATLAS)
exhibited strong emission bands of CN, C$_2$, C$_3$,
and NH$_2$ which are superimposed on
a dust scattering continuum,
typical of cometary spectra in the optical.
The production rates of CN, C$_2$, and C$_3$
derived using the Haser model
and the corresponding C$_2$/CN and C$_3$/CN ratios
suggest that C/2019 Y4 (ATLAS) is a ``typical'' 
Oort cloud comet under the A'Hearn classification, although 
it appears less dusty as revealed by the $Af\rho$ quantities.
Its dust-scattering reflectivity is slightly red,
with a gradient of $\simali$5\% per $10^3\Angstrom$.
We model the reflectivity gradient in terms of porous dust
and find that the red color is accounted for by porous dust.
\end{abstract}



\section{Introduction} \label{sec:intro}

Cometary nuclei are the most primitive objects in the solar system and
have preserved pristine materials from the presolar molecular cloud
and from the early stages of the protosolar nebula. Therefore, the
origin of cometary nuclei is closely linked to the origin of the solar
system. The chemical and physical properties of dust and volatile
materials in cometary nuclei and comae shed light on the nature of the 
primordial interstellar materials present during the solar system 
formation, and provide clues to the processes of incorporation of 
these materials into cometary nuclei, as well as the chemical and 
thermodynamic conditions in the outer solar nebula in which comets 
formed.

In the absence of a direct analysis of cometary nucleus materials, one
often resorts to cometary comae formed by volatiles outgassed from the
nucleus and dust particles dragged out by the expanding gas. Once
formed and stored in the Oort cloud or the Kuiper belt, however,
comets are exposed for $\simali$4.5 billion years to the flux of
Galactic cosmic rays and may develop a substantial ``crust'' of
non-volatile materials \citep[e.g., see][]{1999SSRv...90..269S}. For
short-period comets, such a crust could also result from the solar
radiation during their numerous close approaches to the Sun \citep[e.g., see][]{
1998A&A...338..364L}. In this scenario, only the sub-surface materials 
still remain pristine. Therefore, disintegrating comets provide us an 
unique opportunity to gain access to the pristine, sub-surface materials.

Comet C/2019 Y4 (ATLAS) (hereafter, 19Y4), discovered by the Asteroid 
Terrestrial-impact Last Alert System (ATLAS) at the Mauna Loa Observatory 
on 2019 December 28, is a long-period comet.
According to the JPL orbital solution \#17,
it takes 19Y4 $5895\pm23\yr$ to orbit the Sun
in an elliptical trajectory of eccentricity $e=0.999$,
inclination $i=45.4^{\circ}$, and perihelion distance 
$q=0.253\AU$.\footnote{\url{https://ssd.jpl.nasa.gov/tools/sbdb\_lookup.html\#/?sstr=2019Y4}} 
Its initial light curve exhibited such a steep brightening that, if 
continued, 19Y4 would have been visible to naked eyes when approaching its 
perihelion in 2020 late May.\footnote{\url{http://astro.vanbuitenen.nl/comet/2019Y4}} 
However, the steep brightening terminated
in 2020 mid-March, when 19Y4 started to disintegrate,
as revealed by the nongravitational effect
and a blue color likely caused by the release of
a large amount of volatiles \citep{2020AJ....160...91H}.
The polarization increased dramatically
in 2020 late-March as well, indicating a large increase
in the amount of carbonaceous material
\citep{2020MNRAS.497.1536Z}.
On 2020 April 6, \citet{2020ATel13620....1Y}
and \citet{2020ATel13622....1S} successively
reported an elongated nucleus.
On 2020 April 13, the nucleus of 19Y4
had already split into at least five condensations 
\citep{2020CBET.4751....1}.
Optical spectroscopy on 2020 April 14 and 16
revealed emission bands of CN, C$_2$, C$_3$, and NH$_2$.
The C$_2$/CN and C$_3$/CN production-rate ratios
suggested a ``typical'' Oort cloud comet
under the A’Hearn classification \citep{2021MNRAS.507.5376I}.
Continuing disintegration was reported
by \citet{2020ATel13651....1Y},
and the evolution of the fragment clusters
was investigated in detail
through high angular resolution images
obtained with the {\it Hubble Space Telescope} (HST)
on 2020 April 20 and 23 \citep{2021AJ....162...70Y}.

To explore the properties of the pristine volatiles and dust beneath the refractory crust, we have performed long-slit spectroscopic and broad-band photometric observations of 19Y4 during its disintegration. This paper reports and analyzes the spectroscopic and photometric data and is organized as follows. The observation and data reduction are described in \S\ref{sec:obs} and the results are reported in \S\ref{sec:results}. The gas and dust productions are discussed in \S\ref{sec:gasprod} and \S\ref{sec:dustprod}, respectively. The dust-scattered light is modeled in \S\ref{sec:dustscatt} to infer the properties of the dust. We summarize our major conclusions in \S\ref{sec:summary}.

\section{Observation and Data Reduction} \label{sec:obs}

\subsection{Long-slit Spectroscopy} \label{sec:spec}
We obtained long-slit optical spectra of 19Y4 with the Beijing Faint
Object Spectrograph and Camera (BFOSC) on board the 2.16-Meter
Telescope at Xinglong Station. BFOSC has a $2048\times2048$ pixel$^2$
CCD installed, subtending a $9.36\arcmin\times9.36\arcmin$ field of
view (FOV). The pixel scale is thus $\simali$0.274$\arcsec$
and no binning was performed.
The slit length of 9.4$\arcmin$ was fixed 
along the north-south orientation. We note that the north-south 
orientation is parallactic only when the target is on the meridian. So 
our spectra are somewhat subjected to atmospheric dispersion (see 
\S\ref{sec:results}). A slit width of 1.8$\arcsec$ 
or 2.3$\arcsec$ was chosen, depending on
the seeing at the time of observation.
The G4 grism, covering a wavelength range of
3600--8700$\Angstrom$, was used. 
It has a resolving power of $\sim$265
at [O\,{\sc iii}] $\lambda$4959$\Angstrom$
and $\lambda$5007$\Angstrom$, if combined with a slit
width of 2.3$\arcsec$ \citep{2016PASP..128k5005F}.

\begin{figure}[ht!]
    \centering
    \includegraphics[width=0.5\textwidth]{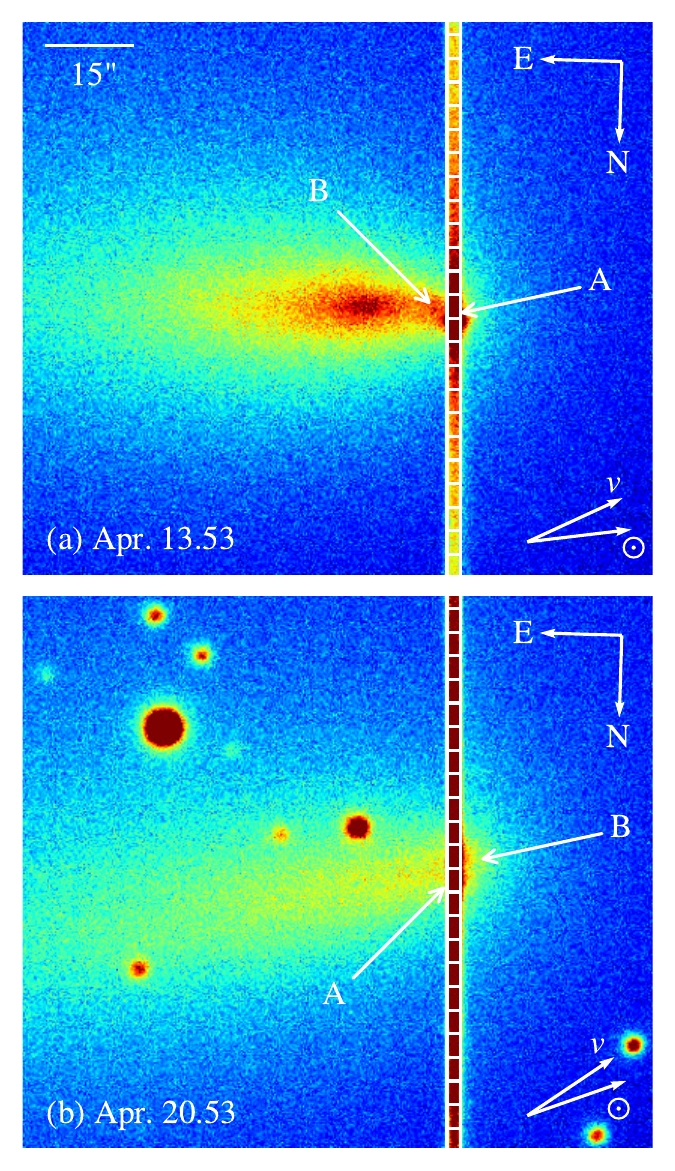}
    \caption{\footnotesize
    Slit views of C/2019 Y4 (ATLAS) on 2020 April 13.53 and 20.53 (UT). The ephemeris positions of the two major fragments A and B (from MPEC 2020-L06 and 2023-J29, respectively) are marked. On April 13.53 (UT), the slit was centered on A. On April 20.53 (UT), the slit was over both A and B. The rectangles, 4.1$\arcsec$ in height and 2.3$\arcsec$ in width, along the slits are sub-apertures. They were used to extract sub-aperture spectra (see \S\ref{sec:results}). The arrows in the upper right corner mark the north and the east directions, and those in the lower right corners mark the velocity and comet-Sun vectors.\label{fig:slitview}} 
\end{figure}

We performed spectroscopic observations of 19Y4 in four nights: 2020
April 6.51, 13.53, 20.53, and 23.58 (UT). Although the disintegration
of the nucleus had already started in mid-March 2020
\citep{2020AJ....160...91H}, no fragment was resolved in our
observation on April 6.51 (UT), and the slit was centered on the
brightest pixels of the elongated nucleus. On April 13.53 (UT), the
slit was centered on the fragment 19Y4-A (see Figure\
\ref{fig:slitview}a). On April 20.53 (UT), fragments 19Y4-A and 19Y4-B
were resolved and they roughly aligned along the north-south
orientation (see Figure\ \ref{fig:slitview}b), therefore, the slit was
over both of them. On April 23.58 (UT), the slit was centered on the
brightest fragment 19Y4-B. Non-sidereal tracking was used.  We alternated between short exposures of 600$\s$ and slit view inspections in order to ensure the guiding accuracy. The iron-argon arc was used as the wavelength calibrator. In addition, we have also obtained the long-slit spectra of several flux standards: Feige 34, He 3, HR 4554, HD 109995, HZ 44 and Feige 98. The spectroscopy log is shown in Table\ \ref{tab:obslog}. Also tabulated in Table\ \ref{tab:obslog} are the kinematic parameters of 19Y4, including its phase angle ($\phi$), heliocentric ($\rh$), and geocentric distances ($\Delta$) as well as their temporal changing rates $\dot{\rh}$ and $\dot{\Delta}$, obtained from the JPL Horizons On-Line Ephemeris System web-interface.\footnote{\url{https://ssd.jpl.nasa.gov/horizons.cgi}}

Following the general data reduction procedures, we reduced the long-slit spectroscopic data with Python. First, we took bias subtraction, corrected for flat fields, and removed cosmic rays. Then, we corrected for the geometric curvature by fitting the curved arc lines with the B-spline functions. Finally, we combined multiple short exposures in each night, and calibrated their wavelengths and fluxes to derive fully reduced 2-D spectra.

\subsection{Broadband Photometry} \label{sec:phot}

BFOSC was also used to perform broadband photometry over 19Y4 on 2020
April 23.55 (UT). The CCD configuration is the same as that used in the long-slit spectroscopy (see \S\ref{sec:spec}). We made three exposures
of 300$\s$ each, with the {\it BVRI} filters in the Johnson-Cousins
system. Again, non-sidereal tracking was used. The photometry log is also shown in Table\ \ref{tab:obslog}.

We developed a Python pipeline to reduce the photometric data. First,
we corrected all the science frames for bias, flat field and cosmic
ray pixels. Then, we smoothed the corrected frames to detect and
center the peaks of the trailed field stars, which were flattened
under non-sidereal tracking. We performed photometry at each
centroid on the corrected frames with rectangular aperture and
annulus to obtain instrumental magnitudes.
Next, we used the centroids of the star trails
to derive plate solutions
via {\it Astrometry.net} \citep{2010AJ....139.1782L},
and the instrumental magnitudes to derive flux zeropoints
by cross-match with the Sloan Digital Sky Survey Data
Release 16 catalog \citep{2022yCat.5154....0A}.
The {\it BVRI} fluxes and errors of the field stars
were constructed from the Sloan {\it gri} fluxes
and errors according to the transformation
relations given in \citet{2008AJ....135..264C}.
The flux calibrated frames were finally combined
according to their filters.

\section{Results} \label{sec:results}

We adopt two apertures, different in width, to extract 1-D spectra: the narrow one, with a single aperture, has a width of $\simali$27.5$\arcsec$ (corresponding to a physical scale of $\simali$2.4\,$\times$\,$10^4\km$); the broad one, divided into 40 sub-apertures, has a width $\simali$165$\arcsec$ in total (corresponding to a physical scale of $\simali$1.2\,$\times$\,$10^5\km$; see Figure\,\ref{fig:slitview}). The narrow aperture is chosen to maximize the signal-to-noise ratio in the dust continuum bands, and the extracted spectra are used to determine the reflectivity gradients (see below). We show in Figure\ \ref{fig:spec} the narrow aperture spectra of 19Y4. The spectra are composed of emission bands from gases and sunlight scattered by dust grains. With the help of an emission line catalog \citep{1996AJ....112.1197B}, we identify the $\Delta\nu=0$ bands of CN, C$_2$, and C$_3$, the $\Delta\nu=+1$ band of C$_2$, and a series of bands of NH$_2$. The sub-aperture spectra are used to both estimate the influence of atmospheric dispersion on the reflectivity gradients (see below) and model the gas and dust production (see \S\ref{sec:gasprod} and \S\ref{sec:dustprod}).

\begin{figure*}[ht!]
    \plotone{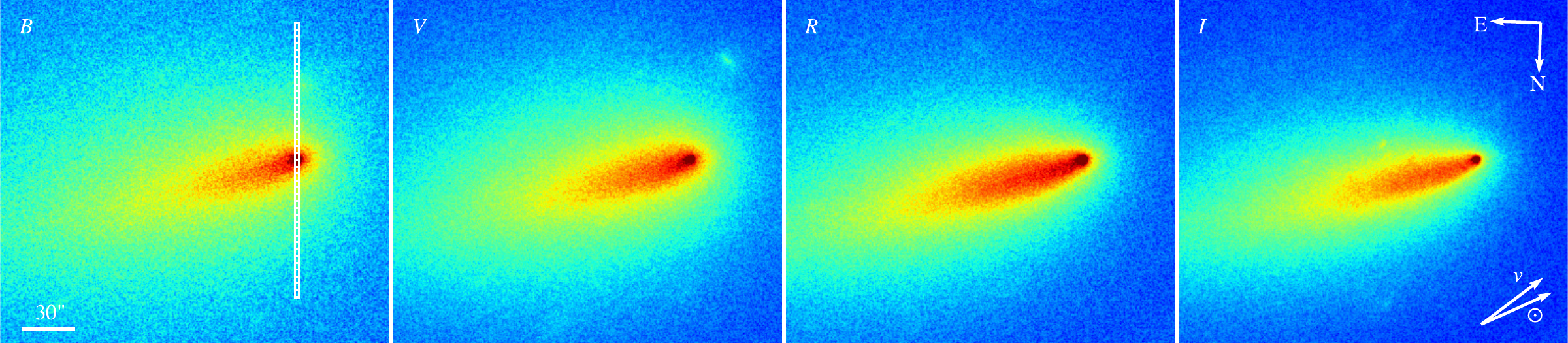}
    \caption{\footnotesize
    {\it BVRI} images of C/2019 Y4 (ATLAS) on 2020 April 23.55 (UT). The coma shows
    a less-extended morphology from {\it B} to {\it I}, indicating the broadband colors vary from inner to outer coma. The white rectangle is a slit-like aperture. It is divided into a series of sub-apertures, 4.1\arcsec in height and 2.3\arcsec in width, to derive $\BV$ and $\VR$ profiles (see \S\ref{sec:results}). The arrows in the upper right corner mark the north and the east directions, and those in the lower right corners mark the velocity and comet-Sun vectors. \label{fig:phot}}
\end{figure*}

\begin{figure*}[ht!]
    \plottwo{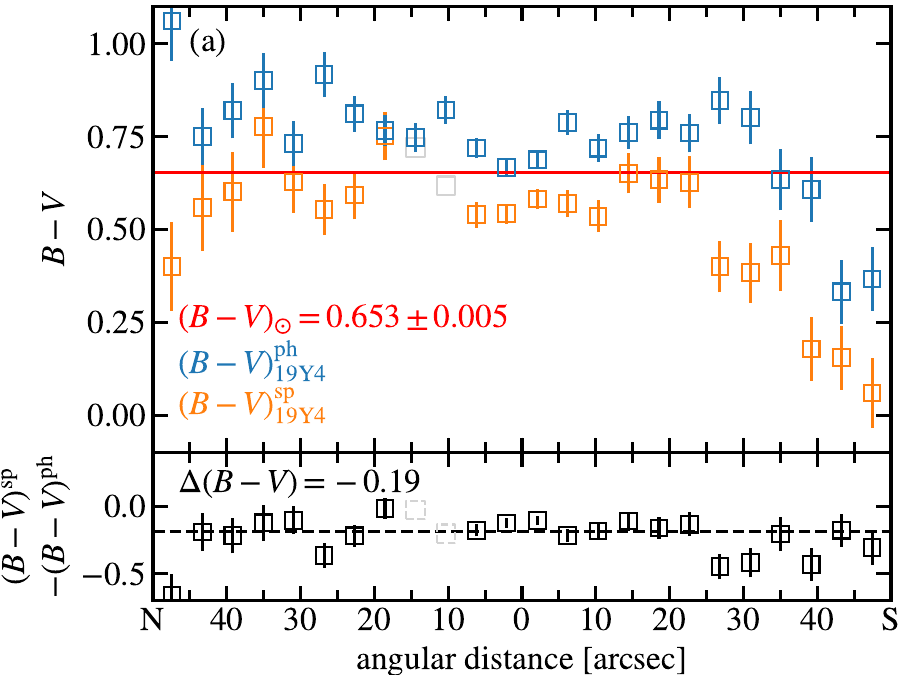}{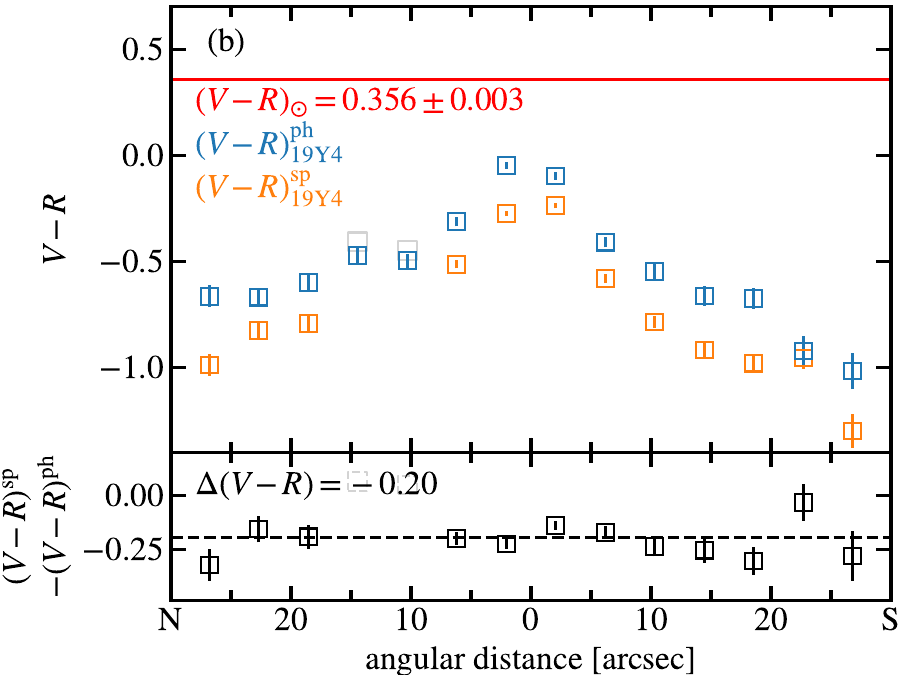}
    \caption{\footnotesize
    Comparison of the $\BV$ (left panel) and $\VR$ (right panel)
    profiles of C/2019 Y4 (ATLAS) derived from photometry (blue open
    squares) and spectroscopy (yellow open squares), both on 2020
    April 23 (UT). Small sections of the spectroscopic profiles are
    contaminated by field stars (grey open squares). Also shown are
    the solar colors (red lines). The spectroscopic profiles
    systematically deviate from the photometric profiles by $-0.19$
    and $-0.20$ in $\BV$ and $\VR$, respectively,
    which are shown as black open squares
    with fitted black dashed lines.
    Note that the error bars only account for
      the propagated Poisson errors. The omitted
      systematic errors can be approximated by
      the standard deviations of $\Delta\BV$
      and $\Delta\VR$, which are $\simali$0.15
      and 0.08\,mag, respectively.\label{fig:color}} 
\end{figure*}

Figure\ \ref{fig:phot} shows the {\it BVRI} images of 19Y4 obtained
with BFOSC on 2020 April 23.55 (UT). As the band goes from {\it B} to
{\it I}, the coma of 19Y4 shows a less-extended morphology, indicating
the broadband colors vary from inner to outer coma. To characterize
the color variation, we derive $\BV$ and $\VR$ profiles of the coma by
performing photometry with a slit-like aperture along the north-south
orientation. The aperture is divided into a series of sub-apertures,
$4.1\arcsec$ in height and $2.3\arcsec$ in width (see Figure\
\ref{fig:phot}), which are the same as that used to extract
sub-aperture spectra. The color profiles are shown in Figure\
\ref{fig:color}. In general, the $\BV$ profile is slightly
redder and the $\VR$ profile is significantly bluer than
the solar colors of $\BV_\odot=0.653\pm0.005$
and $\VR_\odot=0.352\pm0.007$ \citep{2012ApJ...752....5R}.
More specifically, the $\BV$ profile has a local minimum
of 0.68 near the nucleus and becomes redder outwards.
After reaching $\simali$0.9 at $\simali$30$\arcsec$
from the nucleus, it starts to turn blue. 
The $\VR$ profile has a maximum of $-0.07$
near the nucleus and becomes bluer outwards.
As the broad {\it BVR} bands cover multiple
gas emission lines and dust continuum,
such color variation should not be simply
attributed to the intrinsic changes
in the nature of gas or dust. Instead,
it may merely reflects the spread of gases
and dust in the coma.
To be more quantitative, we estimate from
the spectra that the gas emission bands
contribute $\simali$50\% and $\simali$25\%
of the {\it VR} fluxes, respectively.
This suggests that the blue color of $\VR$
does not necessarily indicate
a blue dust continuum of 19Y4.

To characterize the dust color,
we define seven dust continuum bands
free from gas emission and telluric absorption:
5225--5255$\Angstrom$,
5800--5850$\Angstrom$,
6410--6450$\Angstrom$,
6800--6850$\Angstrom$,
7420--7450$\Angstrom$,
7540--7580$\Angstrom$,
7795--7821$\Angstrom$,
and 8000--8050$\Angstrom$,
and derive the reflectivity gradient,
following \citet{1986ApJ...310..937J}.
Observationally, the reflectivity is defined as
\begin{equation}
 S^{\rm obs}_{\lambda}(\rh)=\frac{F^{\rm sca}_{\lambda}(\rh)}{F^{\sun}_{\lambda}(1\AU)} \times \left(\frac{\rh}{\rm AU}\right)^2~~,
\end{equation}
where $F^{\rm sca}_{\lambda}(\rh)$ is the dust scattered continuum of
19Y4 at a heliocentric distance of $\rh$; $F^{\sun}_{\lambda}(1\AU)$
is the solar flux at $\rh=1\AU$. Here we use the solar reference
spectrum from CALSPEC \citep{2014PASP..126..711B}. Since the spectra
obtained at different epochs have different dust continuum levels, the reflectivity is further normalized to $\langle S^{\rm obs}\rangle$, the mean reflectivity in the observed wavelength range. We fit $S^{\rm obs}_{\lambda}/\langle S^{\rm obs}\rangle$ with linear functions and determine the reflectivity gradient, defined as the slope of the best-fit. 

\begin{figure}[ht!]
    \centering
    \includegraphics[width=0.6\textwidth]{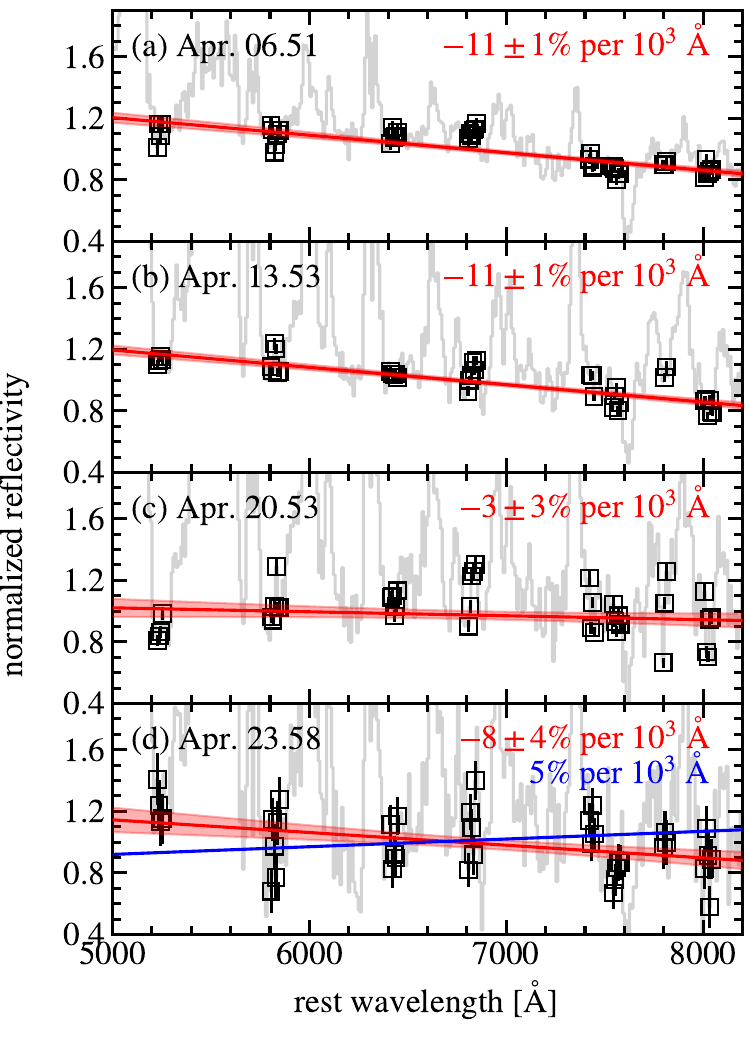}
    \caption{\footnotesize
    The ``observed'' reflectivities ($S^{\rm obs}_{\lambda}/\langle
    S^{\rm obs}\rangle$; gray lines) of C/2019 Y4 (ATLAS) on 2020 April 6.51 (a),
    13.53 (b), 20.53 (c), and 23.58 (UT; d). The gradients are
    determined by fitting those in the dust continuum bands (black
    open squares with error bars)   with linear functions (red lines
    with light red shadows representing the fitting uncertainties). As
    the spectra of C/2019 Y4 (ATLAS) are somewhat subjected to atmospheric dispersion, the gradients need correction. To the first-order approximation, the reflectivity gradient on April 23.58 (UT) is corrected to $\simali$5\% per $10^3\Angstrom$ (blue line; see \S\ref{sec:results} for details). All quantities are normalized to their mean values in the observed wavelength range.\label{fig:ref}}
\end{figure}

As shown in Figure\ \ref{fig:ref}, the reflectivity gradients are
negative at wavelengths 5000--8200$\Angstrom$ for all the four epochs,
indicating that 19Y4 is significantly bluer than the Sun for which, by
definition, the reflectivity gradient is 0$\%$ per $10^3\Angstrom$. It
is also bluer than typical comets for which the reflectivity gradient
is $\simali$5--18\% per $10^{3}\Angstrom$ at wavelengths
3500--6500$\Angstrom$ \citep[see][]{1986ApJ...310..937J}.
We note that \citet{2020AJ....160...91H} also found that 19Y4
at one point had a color that was much bluer than the Sun.
However, as we mentioned in \S\ref{sec:spec}, the spectra of
19Y4 were affected by atmospheric dispersion and so was the color.
Therefore, correction is needed to recover the true reflectivity gradient.


To estimate the influence of atmospheric dispersion on the
reflectivity gradient, we further derive spectroscopic $\BV$ and $\VR$
profiles on April 23.58 (UT) by convolving the sub-aperture spectra
with the the transmission functions of the {\it BVR} filters, and
compare them in Figure\ \ref{fig:color} with the photometric color
profiles on the same date. If the spectra were free from atmospheric
dispersion, there should not be any difference between the
spectroscopic and photometric color profiles, as the apertures used to
derive the color profiles are the same. Nevertheless, as shown in
Figure\ \ref{fig:color}, the spectroscopic $\BV$ and $\VR$ colors are
systematically bluer, indicating the influence of the atmospheric
dispersion on the colors. To the first-order approximation, we assume
that the color deviation due to the atmospheric dispersion varies
linearly with the wavelength, and determine a linear correction
function by minimizing the systematic differences in the color
profiles. We note that such a correction is only applied
to the spectrum obtained on 2020 April 23.58 (UT) and,
as shown in Figure~\ref{fig:ref}, 
the recovered reflectivity gradient is 
$\simali$5\% per $10^3\Angstrom$
at wavelengths 5000--8200$\Angstrom$. 
The results are shown in Figure~\ref{fig:ref}
and will be used to constrain the dust properties
in \S\ref{sec:dustscatt}.

\section{Discussion} \label{sec:discussion}

\subsection{Gas Production} \label{sec:gasprod}
\begin{figure*}[ht!]
    \plotone{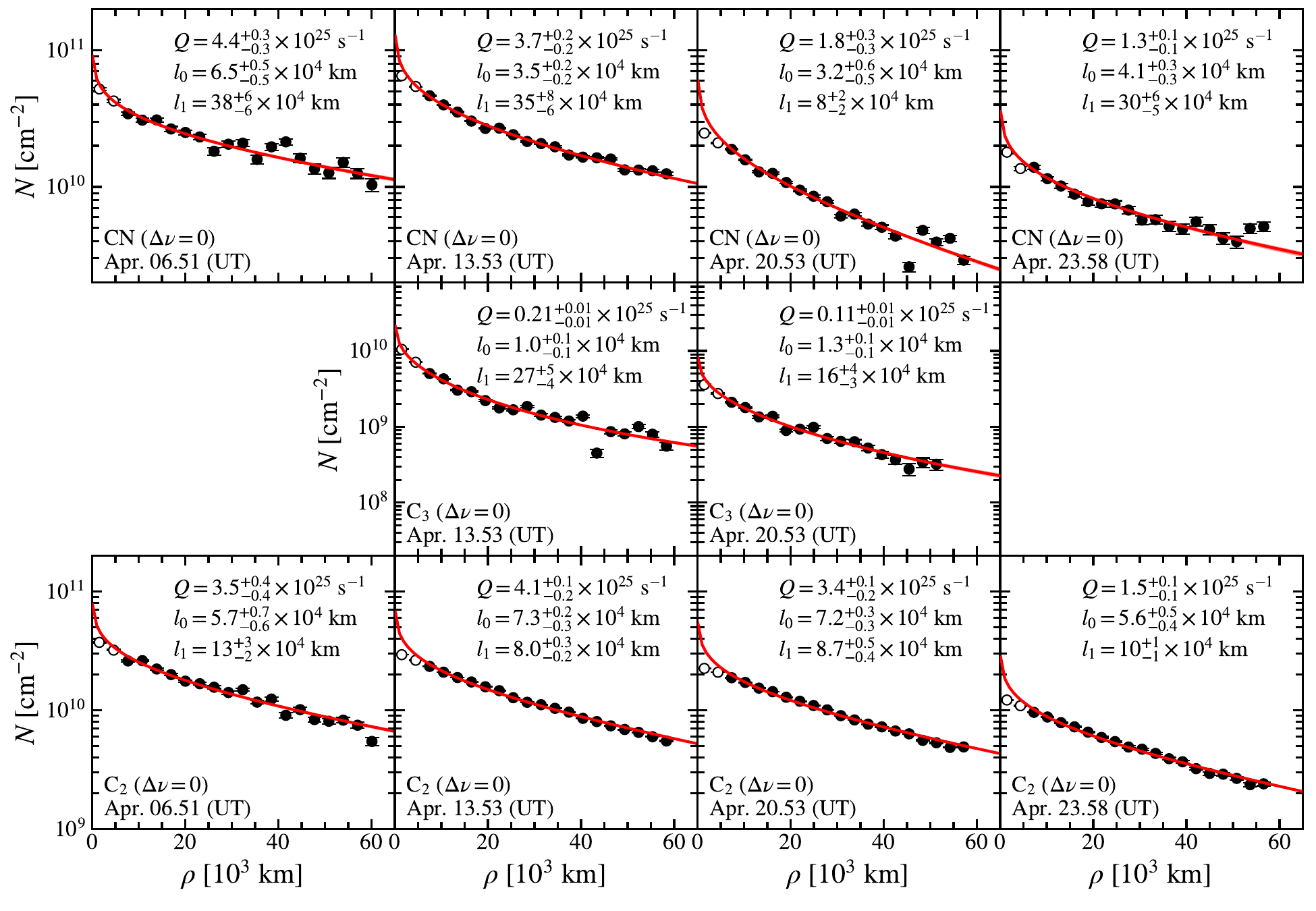}
    \caption{\footnotesize
    Comparison of the ``observed'' column density profiles (filled black circles with error bars) of CN ($\Delta\nu=0$; upper panels), C$_3$ ($\Delta\nu=0$; middle panels), and C$_2$($\Delta\nu=0$; bottom panels) with model fits (solid red lines with light red shadows representing the model uncertainties). In each panel the model parameters are labelled. The open black circles are not considered for modeling (see \S\ref{sec:gasprod} for details). \label{fig:gasprod}}
\end{figure*}

The gas production rate and scale length of a given volatile species are derived by modeling the ``observed'' column density profile, which, for species $j$ at a projected distance $\rho$, is
\begin{equation}
    N^{\rm obs}_j=\frac{4\pi I_j(\rho)}{g_j}~~,
\end{equation}
where $I_j(\rho)$ is the continuum-subtracted radiance of the species
measured at $\rho$ and integrated over a bandpass, and $g_j$ is the
fluorescence efficiency of the species averaged over the band
\citep{2011Icar..213..280L}. We average $N^{\rm obs}_j$ over the two
pairs of the sub-apertures symmetric about the nucleus. In this work,
we derive the ``observed'' column density profiles of CN, C$_3$ and
C$_2$ through their $\Delta\nu=0$ bands. The bandpasses and
fluorescence efficiencies are taken from \citet{2011Icar..213..280L}
except for CN ($\Delta\nu=0$), the fluorescence efficiency of
\citet{2010AJ....140..973S} is adopted for which the Swings effect
\citep{1941LicOB..19..131S} is taken into account.
We show in Figure\ \ref{fig:gasprod}
the ``observed" column density profiles
of CN, C$_3$, and C$_2$.

In the literature, cometary gas production rates, scale lengths and
their dependencies on $\rh$ have been carefully investigated in
numerous statistical works \citep[e.g.,
see][]{1995Icar..118..223A,1996ApJ...459..729F,2011Icar..213..280L}. We
apply those results as prior knowledge for Bayesian inference
studies. To this end, we assume a log-uniform prior for the gas
production rate $Q$ and normal priors for both primary ($l_0$) and
product ($l_1$) scale lengths, i.e., $\log
Q\sim\mathcal{U}\left(22,28\right)$,
$l_0\sim\mathcal{N}(\hat{l}_0,\hat{\sigma}^2_{l_0})$ and
$l_1\sim\mathcal{N}(\hat{l}_1,\hat{\sigma}^2_{l_1})$, where
$\hat{l}_0,\hat{\sigma}_{l_0}$ and $\hat{l}_1,\hat{\sigma}_{l_1}$ are
taken from \citet{2011Icar..213..280L}. Here $\mathcal{U}$ and
$\mathcal{N}$ are probability functions. With $\log
Q\sim\mathcal{U}\left(22,28\right)$, we assume that the production
rate $Q$ is uniformly distributed between $10^{22}\s^{-1}$ and
$10^{28}\s^{-1}$ in the log space. As we do not have any knowledge
about $Q$, we should assume such a non-informative prior. $l_0$ and
$l_1$ are scale lengths for both of which we assume Gaussian priors,
as their values have been well investigated
in previous statistical studies.
The Gaussian parameters $\hat{l}_0,\hat{\sigma}_{l_0}$
and $\hat{l}_1,\hat{\sigma}_{l_1}$
(i.e., mean and standard deviation)
are taken from \citet{2011Icar..213..280L}.
The use of Bayesian inference is justified
later in this section. The corresponding prior
probability density function is denoted as
$p\left(Q,l_0,l_1\right)$. From Bayes' theorem,
the posterior density is proportional to the prior
multiplied by a likelihood function, i.e.,
\begin{equation}
    p\left(Q,l_0,l_1\middle|N^{\rm obs};\sigma_{N^{\rm obs}},\rho\right)\propto p\left(N^{\rm obs}\middle|Q,l_0,l_1;\sigma_{N^{\rm obs}},\rho\right)\times p\left(Q,l_0,l_1\right)~~,
\end{equation}
where $\sigma_{N^{\rm obs}}$ is the uncertainty of $N^{\rm obs}$. If $\sigma_{N^{\rm obs}}$ is assumed to be Gaussian, the likelihood function can be defined as
\begin{equation}
    p\left(N^{\rm obs}\middle|Q,l_0,l_1;\sigma_{N^{\rm obs}},\rho\right)=
    \prod_i \frac{1}{\sqrt{2\pi}\sigma_{N^{\rm obs}_i}}\exp\left[-\frac{\left(N^{\rm mod}_i-N^{\rm obs}_i\right)^2}{2\sigma^2_{N^{\rm obs}_i}}\right]~~,
\end{equation}
where $N^{\rm mod}_i$ is the modeled column density in the $i$-th
sub-aperture. With a set of $(Q,l_0,l_1)$ given, $N^{\rm mod}_i$ can
be derived from eqs.\,(11)--(13) in \citet{2011Icar..213..280L}. The
joint posterior distributions are then sampled using the Markov chain
Monte Carlo (MCMC) method. The $50\%$ (median) and $50\pm34\%$ percentiles in the marginalized distributions of $Q$, $l_0$ and $l_1$ are taken as the best-fit values and the uncertainties, respectively. Note that, in the fitting, $N^{\rm obs}_i$ with $\rho_i<5\times10^3\km$ are excluded, considering that the gas distribution in the inner coma may not be spherically symmetric due to the disintegration and also that ``holes'' may be present in the profiles of C$_2$ \citep[see][]{2011Icar..213..280L}. 

The best-fits to the ``observed'' column density profiles are shown in
Figure\ \ref{fig:gasprod}, with the best-fit model parameters and
uncertainties listed in Table\ \ref{tab:gasprod}. The fits to C$_3$
($\Delta\nu=0$) on 2020 April 6.51 and 23.58 (UT) fail
due to the low signal-to-noise ratio of the observational data,
and are thus not shown.
Typically, to simultaneously determine both $l_0$ and $l_1$,
the ``observed'' profiles are required to extend further than $l_1$
\citep{2005P&SS...53.1243F}. However, this is not the case here. If
the widely adopted $\chi^2$-minimizing method is used,
several $\left(l_0,l_1\right)$ pairs could lead to equally good
fits \citep{2005P&SS...53.1243F,2011Icar..213..280L}.
Therefore, it is the Bayesian model or, more specifically,
the prior that helps to find unique best-fits. 

Since G4 does not cover any strong ${\rm OH}$ bands, we take $Q({\rm
  CN})$ as an indicator of the gas production rate of 19Y4. We show in
Figure\ \ref{fig:prod}a the dependency of $Q({\rm CN})$ on $\rh$. In
general, the gas production rate in 19Y4 is ``typical'' in the sense
that both the production rate and its dependency on $\rh$ agree with that of the 26 comets reported in \citet{2011Icar..213..280L}. 

We estimate a mean ratio of $\log\left[Q({\rm C_2})/Q({\rm CN})\right]\approx0.2$ and $\log\left[Q({\rm C_3})/Q({\rm CN})\right]\approx-1.2$ at $\rh=1\AU$, if we assume $\log\left[Q({\rm C_2})/Q({\rm CN})\right]$ and $\log\left[Q({\rm C_3})/Q({\rm CN})\right]$ follow a power-law dependence on $\rh$ and take the power-law slopes of \citet{2011Icar..213..280L}. The C$_2$/CN and C$_3$/CN production rate ratios estimated for 19Y4 suggest a ``typical'' comet instead of a ``depleted'' one under the classification of \citet{1995Icar..118..223A}, consistent with that found by \citet{2021MNRAS.507.5376I}.

\subsection{Dust Production} \label{sec:dustprod}
Initially introduced in \citet{1984AJ.....89..579A}, the so-called $Af\rho$ quantities are often used to describe cometary dust production, where $A$ is the dust reflectivity (i.e., albedo), $f$ is the dust filling factor and $\rho$ is the projected linear radius of the FOV at the comet as discussed in \S\ref{sec:gasprod}. The $Af\rho$ quantity within a circular aperture with a projected radius of $\rho$ is determined directly from observation as

\begin{equation}
    Af\rho=\frac{(2\Delta)^2}{\rho}\left(\frac{\rh}{\rm AU}\right)^2\frac{F_{\rm c}}{F_{\sun}(1\AU)}\frac{1}{\Phi_{\rm HM}(\phi)}~~,
\end{equation}
where $F_{\rm c}$ is the cometary flux in a chosen dust continuum band, $F_{\sun}(1\AU)$ is the solar flux at $\rh=1\AU$ in the same band, and $\Phi_{\rm HM}(\phi)$ is the normalized Halley-Marcus phase function \citep{2011AJ....141..177S}. In most cases, $F_{\rm c}$ is obtained from narrow-band photometry. But it can also be constructed from long-slit spectroscopy by considering a geometric correction function

\begin{equation}
    F_{\rm c}=\sum_{i}G(\rho_i)F(\rho_i)~~,
\end{equation}
where $\rho_i$ is the projection distance from the $i$-th sub-aperture to the comet nucleus, and $G(\rho_i)$ is the geometric correction function \citep{2011Icar..213..280L}. The summation is over all the sub-apertures within $\rho$ (i.e., $\rho_i<\rho$). In this work, we calculate $Af\rho$ from the constructed $F_{\rm c}$ in the $\lambda4850\Angstrom$ and $\lambda5240\Angstrom$ continuum bands refined in \citet{2011Icar..213..280L} with an aperture radius of 10000$\km$. The results are tabulated in Table~\ref{tab:Afrho}.
We should admit that our method of reconstructing
$Af\rho$ from long-slit spectra which only sample
a strip along certain part of the coma has a drawback
of not accounting for the coma’s asymmetry,
even more so for a disintegrating comet.
However, detailed analysis of dust production
during disintegration requires
high resolution, spatially resolved photometry
and detailed modeling, which is beyond the scope of
this work. Nevertheless, we note that our $Af\rho$ values
closely agree with that from narrow-band photometry
measured with TRAPPIST (Y.~Moulane, private conversation).

\begin{figure}[ht!]
    \centering
    \includegraphics[width=0.6\textwidth]{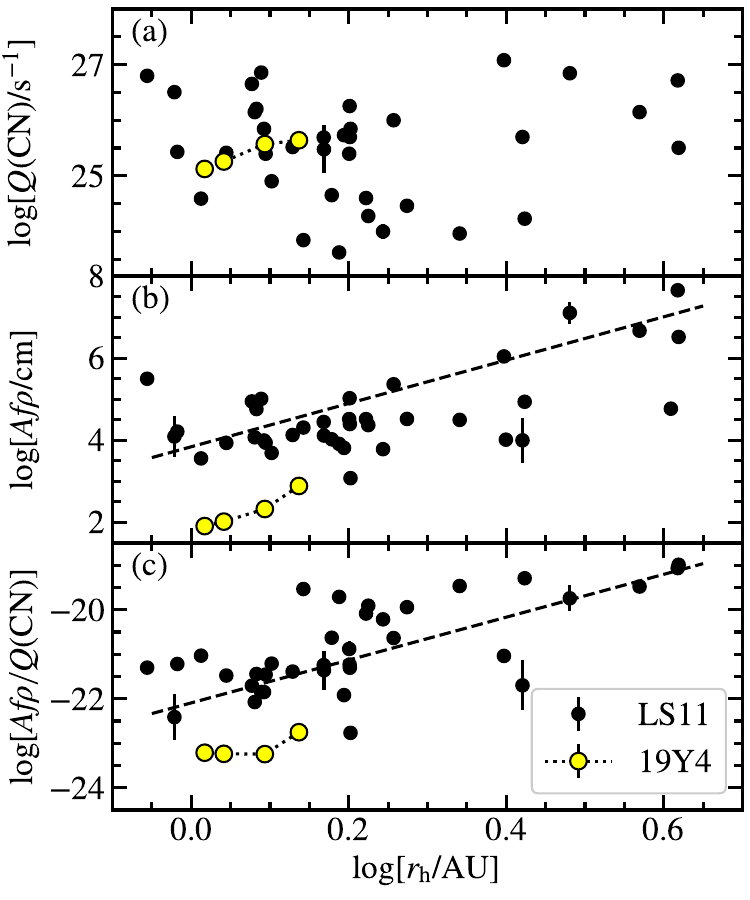}
    \caption{\footnotesize
    Top panel (a): Variation of the CN production rate Q(CN) of C/2019 Y4 (ATLAS) (yellow circles) with heliocentric distance $\rh$. Also shown are the CN production rates of the \citet{2011Icar..213..280L} sample of 26 comets (black circles). Middle panel (b): Same as (a) but for $Af\rho$. Also shown is a linear fit to the $\rh$-dependence of $Af\rho$ for the \citet{2011Icar..213..280L} sample: $d(\log Af\rho)/d(\rh)=5.3\pm0.6$ (black dashed line). Bottom panel: Same as (a) but for $Af\rho/Q({\rm CN})$. The power-law has an index of $4.6\pm0.8$. \label{fig:prod}}
\end{figure}

We obtain a mean $Af\rho$ for 19Y4 by averaging over that in the
$\lambda4850\Angstrom$ and $\lambda5240\Angstrom$ continuum bands and
examine the dependency of $Af\rho$ on $\rh$ in Figure\
\ref{fig:prod}b. For comparison, we also show in Figure\
\ref{fig:prod}b the $Af\rho$ quantities of the 26 comets studied in
\citet{2011Icar..213..280L}. While 19Y4 appears to be less dusty, it
exhibits a $\rh$-dependence more or less resembling that of the
\citet{2011Icar..213..280L} samples for which a power-law with an
index $5.3\pm0.6$ is suggested, even though most comets
in the sample of \citet{2011Icar..213..280L}
did not disintegrate when observed.

\subsection{Scattered Sunlight} \label{sec:dustscatt}

\begin{figure*}[ht!]
    \plotone{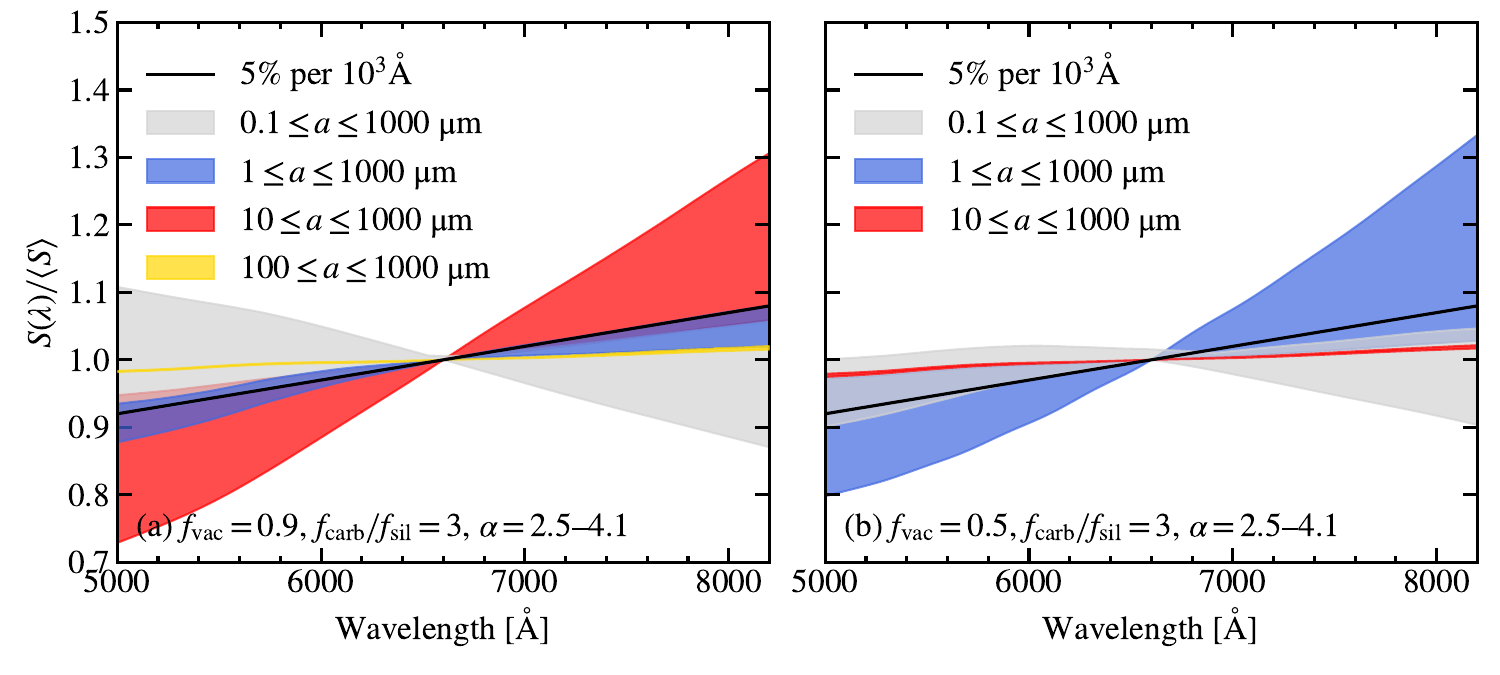}
    \caption{\footnotesize
    Comparison of the dust models with the recovered reflectivity gradient (black line). For the dust, two cases are considered for the porosity (i.e., the fractional volumn of vacuum): $f_{\rm vac}=0.9$ (left panel) and $f_{\rm vac}=0.5$ (right panel). The volume mixing ratio of amorphous silicate and amorphous carbon is fixed to $f_{\rm carb}/f_{\rm sil}=3$. For the size distribution, four size ranges are considered, different in the lower cutoff, i.e., $\amin=0.1$ (gray shadow), 1 (blue shadow), 10 (red shadow), and 100$\um$ (yellow shadow). The upper cutoff ($\amax$) is fixed to 1000$\um$. Given a size range, the power-law index $\alpha$ is varied from 2.5 to 4.1. The recovered reflectivity gradient ($\simali$5\% per $10^3\Angstrom$) is barely explained by both $1\le a\le1000\um$ and $10\le a\le1000\um$ when $f_{\rm vac}=0.9$, while well explained by the size range $1\le a\le1000\um$ when $f_{\rm vac}=0.5$. For both cases, larger or smaller $\amin$ leads to flatter gradient (or ``gray'' in color).\label{fig:dustscatt}}
\end{figure*}

The dust-scattered light observationally determined in
\S\ref{sec:results} can be used to infer the dust size, compositonal,
and structural properties. To reproduce the normalized reflectivity
$S^{\rm obs}_{\lambda}/\langle S^{\rm obs}\rangle$ (see
\S\ref{sec:results} and Figure\ \ref{fig:dustscatt}), we adopt the
cometary dust model of \citet{1998ApJ...498L..83L}, which
assumes cometary dust grains to be porous aggregates of small
astronomical silicates, amorphous carbon, and vacuum. We consider two
cases for the porosity (i.e., the fractional volumn of vacuum):
$f_{\rm{vac}}=0.9$ and $f_{\rm{vac}}=0.5$,
which are roughly the upper and lower limits
for cometary dust grains
\citep[see][]{1990ApJ...361..260G,1998A&A...330..375G}.
For porous dust resulting from coagulational growth,
one would expect a porosity higher than $f_{\rm vac}=0.5$
\citep[see][]{2008ARA&A..46...21B}.
On the other hand, polarimetric measurements suggest that,
for most of the time, amorphous silicates in the coma of 19Y4
account for a fractional volumn ($f_{\rm sil}$)
of 0.17--0.28 \citep{2020MNRAS.497.1536Z}.
Therefore, we assume a bulk volume mixing ratio
of $f_{\rm carb}/f_{\rm sil}=3$ for our dust model.
The refractive indices of amorphous silicates
and amorphous carbon are taken, respectively,
from \citet{1984ApJ...285...89D}
and \citet{1991ApJ...377..526R}.


We use Mie theory combined with the Bruggman effective medium theory \citep{1983asls.book.....B} to calculate $C_{\rm sca}(\lambda, a)$ and $g(\lambda,a)$, the scattering cross sections and asymmetry factors of spherical porous aggregates of radii $a$ at wavelength $\lambda$. Here the aggregate size $a$ refers to the radius of the sphere encompassing the entire aggregate (we assume that all grains are spherical in shape). We assume a power-law dust size distribution, $dn/da\propto a^{-\alpha}$, over the size range $\amin\le a\le \amax$. The model reflectivity, $S^{\rm mod}(\lambda)$, is calculated as

\begin{equation}
    S^{\rm mod}(\lambda)\propto\int^{\amax}_{\amin} C_{\rm sca}(\lambda, a)\Phi_{\rm HG}\left[g(\lambda,a);\theta\right]\frac{dn}{da} da~~,
\end{equation}
where $\theta$, defined as $\pi-\phi$, is the scattering angle, and
$\Phi_{\rm HG}\left[g(\lambda,a);\theta\right]$ is the
Henyey-Greenstein phase function \citep{1941ApJ....93...70H}. Note
that the right-hand side differs from the absolute reflectivity by
some scaling factor. However, similar to $S^{\rm obs}(\lambda)$,
$S^{\rm mod}(\lambda)$ is normalized with respect to the averaged
value over the defined continuum wavelengths to produce $S^{\rm
  mod}_{\lambda}/\langle S^{\rm mod}\rangle$ during which the scaling
factor is cancelled out.
It is apparent that the model reflectivity
$S^{\rm mod}(\lambda)$ contains information
about the dust size, composition and morphology
since the scattering cross section
$C_{\rm sca}(\lambda, a)$ and the asymmetry factor
$g(\lambda,a)$ are dependent on the dust size $a$,
composition (through the refractive index),
and morphology (e.g., porosity).
  
We calculate normalized reflectivities with different size
distributions. Four size ranges are considered, different in the lower
cutoff, i.e., $\amin=0.1$, 1, 10, and 100$\um$. The upper cutoff
($\amax$) is fixed to 1000$\um$.
The resulting reflectivity gradient
is not sensitive to the choice of $\amax$
since the scattering in the optical
by such mm-sized or larger grains
is gray \citep[e.g., see][]{li2008optical}.
Given a size range, we vary $\alpha$ from 2.5 to 4.1, the flattest and the steepest power-law for cometary dust grains from space measurements \citep{1991ASSL..167.1043M,2009Icar..199..129L,2010M&PS...45.1409P}. The parameters are tabulated in Table\ \ref{tab:modpara}.

We compare the model reflectivities in Figure~\ref{fig:dustscatt}
with the reflectivity derived from observations in
\S\ref{sec:results}. Models with same $\amin$ but different $\alpha$
are grouped and shown as shadows of different colors.
Figure~\ref{fig:dustscatt} shows that
the ``observed'' reflectivity gradient
(i.e., $5\%$ per $10^3\Angstrom$) can be explained
by different combinations of dust parameters.
For highly porous dust with $f_{\rm vac}=0.9$, 
models with $\amin=1\mum$ and $\amin=10\mum$
are able to reproduce the ``observed'' reflectivity gradient.
For dust with a lower porosity of $f_{\rm vac}=0.5$,
while models with $\amin=10\mum$ result in
reflectivity gradients too flat to be comparable
with observed, the ``observed'' reflectivity gradient
can be reproduced by dust with $\amin=1\mum$.
Apparently, although the limited observational constraint
does not allow us to uniquely determine the dust parameters,
the porous dust model is capable of accounting
for the ``observed'' reflectivity gradient
with appropriate combination of parameters.
In general, the reflectivity gradient
$S^{\rm mod}(\lambda)$ decreases
as the dust becomes more porous
or $\amin$ becomes smaller.
The variation of $S^{\rm mod}(\lambda)$
with $\alpha$, the power index of the dust
size distribution, depends on the choice
of $\amin$. For $\amin<1\mum$,
$S^{\rm mod}(\lambda)$ becomes bluer
as $\alpha$ increases, due to the increasing
role played by submicron-sized grains
which scatter more at shorter wavelengths.
By increasing $\amin$ to several microns
(and larger), $S^{\rm mod}(\lambda)$ becomes
redder as $\alpha$ increases since these
micron-sized grains scatter more effectively
at longer wavelengths.
With $\amin$ reaching several tens of microns,
$S^{\rm mod}(\lambda)$ becomes gray
and does not vary with $\alpha$
since these grains are in the geometrical
optics limit and their scattering does not
vary much with wavelength.
Apparently, $\amin$ plays a more important role
than $\alpha$ in determining $S^{\rm mod}(\lambda)$.

\section{Summary} \label{sec:summary}
We presented long-slit optical spectra
at $\simali$3600--8700$\Angstrom$
and multi-band $BVRI$ images
of the disintegrating comet 19Y4,
from which we derived the gas and dust production rates
and modeled the dust-scattering reflectivity.
Our principal results are as follows:
\begin{enumerate}
\item All of the spectra taken during the disintegration
          showed strong gas emission. The production rates
          of CN, C$_2$, and C$_3$ and the corresponding
           C$_2$/CN and C$_3$/CN production-rate ratios
           were derived. Both the gas production rates and
           their dependencies on $\rh$ are consistent with 
           that of ``typical'' Oort cloud comets,
           not that of carbon-chain ``depleted'' Kuiper Belt comets. 
\item The dust-scattering reflectivity of 19Y4 was
             somewhat red, with a gradient of $\simali$5\%
             per $10^3\Angstrom$. The reflectivity gradient
             was modeled in terms of porous dust.
             The red color is accounted for by porous dust.
\end{enumerate}

\acknowledgments
We thank the anonymous referees
for their valuable comments and suggestions
which improved the quality and presentation
of this work.
RNZ thanks Dan Yu for her love and companionship over the past eight years, and dedicates this article to their marriage. We acknowledge the support of the staff of the Xinglong 2.16m telescope. This work was partially Supported by the Open Project Program of the CAS Key Laboratory of Optical Astronomy, National Astronomical Observatories, Chinese Academy of Sciences. RNZ and JFL are supported by NSFC through grant numbers 11988101 and 11933004. 
LW is supported by NSFC through grant number U2031144. JFL acknowledges the support from the New Cornerstone Science Foundation through the New Cornerstone Investigator Program and the XPLORER PRIZE.

%

\vspace{5mm}
\facilities{Beijing:2.16m (BFOSC)}

\software{astropy \citep{2013A&A...558A..33A},
          ccdproc \citep{2017MNRAS.469S.712B},
          photutils \citep{Bradley_2019_2533376},
          astroquery \citep{2019AJ....157...98G},
          specutils \citep{2019ascl.soft02012A},
          sbpy \citep{2019JOSS....4.1426M},
          lmfit \citep{2014zndo.....11813N},
          emcee \citep{emcee},
          }

\begin{deluxetable*}{cccccccccccc}
    \tablenum{1}
    \tablecaption{Log of BFOSC observations.\label{tab:obslog}}
    \tablewidth{0pt}
    \tablehead{
        \colhead{Date of 2020} & \colhead{Object} & \colhead{Grism/Filter} & \colhead{Slit Width} & \colhead{Exposure} & \colhead{Airmass} & \colhead{$\rh$\tablenotemark{a}} & \colhead{$\dot{\rh}$} & \colhead{$\Delta$\tablenotemark{b}} & \colhead{$\dot{\Delta}$} & \colhead{$\phi$\tablenotemark{c}} \\
        \colhead{(UT)} & \nocolhead{} & \nocolhead{} & \colhead{($\arcsec$)} & \colhead{(s)} & \nocolhead{} & \colhead{(AU)} & \colhead{(${\rm{km\ s^{-1}}}$)} & \colhead{(AU)} & \colhead{(${\rm{km\ s^{-1}}}$)} & \colhead{($^\circ$)}
    }
    \decimals
    \startdata
    Apr.  6.51 & 19Y4 & \multirow{2}{*}{G4} & \multirow{2}{*}{1.8} & $6\times600$ & 1.215 & 1.37 & -32.4 & 1.03 & -5.55 & 46.6\\
    Apr.  6.54 & Feige 34 & & & 600 & 1.012 & & & & & \\\hline
    Apr. 13.53 & 19Y4 & \multirow{2}{*}{G4} & \multirow{2}{*}{2.3} & $6\times600$ & 1.279 & 1.24 & -33.7 & 1.00 & -6.03 & 51.8\\
    Apr. 13.56 & He 3 & & & 600 & 1.389 & & & & & \\\hline
    Apr. 20.53 & 19Y4 & \multirow{3}{*}{G4} & \multirow{3}{*}{2.3} & $4\times600$ & 1.404 & 1.10 & -35.4 & 0.98 & -7.33 & 57.2\\
    Apr. 20.56 & HR 4554 & & & 5 & 1.037 & & & & & \\
    Apr. 20.59 & HD 109995 & & & 120 & 1.014 & & & & & \\\hline
    Apr. 23.55 & 19Y4 & {\it BVRI} & & $3\times300$ & 1.526 & 1.04 & -36.1 & 0.97 & -8.10 & 59.9\\\hline
    Apr. 23.58 & 19Y4 & \multirow{4}{*}{G4} & \multirow{4}{*}{1.8} & $1\times600$ & 1.763 & 1.04 & -36.1 & 0.97 & -8.10 & 59.9\\
    Apr. 23.59 & HD 109995 & & & 30 & 1.005 & & & & & \\
    Apr. 23.67 & HZ 44 & & & 360 & 1.012 & & & & & \\
    Apr. 23.84 & Feige 98 & & & 600 & 1.412 & & & & & \\
    \enddata
    \tablenotetext{a}{Heliocentric distance}
    \tablenotetext{b}{Geocentric distance}
    \tablenotetext{c}{Phase angle}
\end{deluxetable*}

\begin{deluxetable*}{cccccccccccc}
    \tablenum{2}
    \tablecaption{Gas production rates and scale lengths.\tablenotemark{a}\label{tab:gasprod}}
    \tablewidth{0pt}
    \tablehead{
        \colhead{Date of 2020} & \colhead{$\rh$} & \multicolumn3c{CN} & \multicolumn3c{C$_3$} & \multicolumn3c{C$_2$} & \colhead{Ref.\tablenotemark{b}}\\
        \cline{3-11}
        \nocolhead{} & \nocolhead{} & \colhead{$Q$} & \colhead{$l_0$} & \colhead{$l_1$} & \colhead{$Q$} & \colhead{$l_0$} & \colhead{$l_1$} & \colhead{$Q$} & \colhead{$l_0$} & \colhead{$l_1$}\\
        \colhead{(UT)} & \colhead{(AU)} & \colhead{($10^{25}\,{\rm s^{-1}}$)} & \colhead{($10^4\km$)} & \colhead{($10^4\km$)} & \colhead{($10^{25}\,{\rm s^{-1}}$)} & \colhead{($10^4\km$)} & \colhead{($10^4\km$)} & \colhead{($10^{25}\,{\rm s^{-1}}$)} & \colhead{($10^4\km$)} & \colhead{($10^4\km$)}
    }
    \decimals
    \startdata
    Apr.  6.51 & 1.37 & $4.4^{+0.3}_{-0.3}$ & $6.5^{+0.5}_{-0.5}$ & $38^{+6}_{-6}$ & $-$                  & $-$               & $-$          & $3.5^{+0.4}_{-0.4}$ & $5.7^{+0.7}_{-0.6}$ & $13^{+3}_{-2}$ & 1 \\
    Apr. 13.53 & 1.24 & $3.7^{+0.2}_{-0.2}$ & $3.5^{+0.2}_{-0.2}$ & $35^{+8}_{-6}$ & $0.21^{+0.01}_{-0.01}$ & $1.0^{+0.1}_{-0.1}$ & $27^{+5}_{-4}$ & $4.1^{+0.1}_{-0.2}$ & $7.3^{+0.2}_{-0.3}$ & $8.0^{+0.3}_{-0.2}$ & 1 \\
    Apr. 14.74 & 1.21 & $<$1.05 & 3.2 & 43 & $<$0.015 & 0.15 & 8.8 & $<$0.83 & 2.4 & 16 & 2\\
    Apr. 16.77 & 1.17 & $<$1.12 & 3.0 & 41 & $<$0.015 & 0.14 & 8.2 & $<$0.79 & 2.2 & 15 & 2\\
    Apr. 20.53 & 1.10 & $1.8^{+0.3}_{-0.3}$ & $3.2^{+0.6}_{-0.5}$ & $8^{+2}_{-2}$& $0.11^{+0.01}_{-0.01}$ & $1.3^{+0.1}_{-0.1}$ & $16^{+4}_{-3}$ & $3.4^{+0.1}_{-0.2}$ & $7.2^{+0.3}_{-0.3}$ & $8.7^{+0.5}_{-0.4}$ & 1 \\
    Apr. 23.58 & 1.04 & $1.3^{+0.1}_{-0.1}$ & $4.1^{+0.3}_{-0.3}$ & $30^{+6}_{-5}$ & $-$                  & $-$               & $-$          & $1.5^{+0.1}_{-0.1}$ & $5.6^{+0.5}_{-0.4}$ & $10^{+1}_{-1}$ & 1 \\
    \enddata
    \tablenotetext{a}{The production rates of \citet{2021MNRAS.507.5376I}
      were merely upper limits. Also, when \citet{2021MNRAS.507.5376I}
      derived the production rates from the Haser model,
      they adopted a set of scale lengths
      which are different from ours.
      Note that our rates are consistent with
      that derived from the TRAPPIST observations 
      (Y.~Moulane, private conversation).}
    \tablenotetext{b}{Reference: (1) This work; (2) \citet{2021MNRAS.507.5376I}}
\end{deluxetable*}

\begin{deluxetable}{cccc}
    \tablenum{3}
    \tablecaption{$Af\rho$ within $\rho=10000\km$ in $\lambda4850\Angstrom$ and $\lambda5240\Angstrom$ bands. \label{tab:Afrho}}
    \tablewidth{0pt}
    \tablehead{
        \colhead{Date of 2020} & $\rh$ & \multicolumn2c{$Af\rho$} \\
        \cline{3-4}
        \nocolhead{} & \nocolhead{} & \colhead{$\lambda4850\Angstrom$} & \colhead{$\lambda5240\Angstrom$} \\
        \colhead{(UT)} & \colhead{(AU)} & \colhead{(cm)} & \colhead{(cm)}
    }
    \decimals
    \startdata
    Apr.  6.51 & 1.37 & 972$\pm$10 & 753$\pm$9\\
    Apr. 13.53 & 1.24 & 288$\pm$1  & 199$\pm$1\\
    Apr. 20.53 & 1.10 & 155$\pm$2  &  97$\pm$1\\
    Apr. 23.58 & 1.04 & 145$\pm$3  &  73$\pm$3\\
    \enddata
\end{deluxetable}

\begin{deluxetable}{ccccc}
    \tablenum{4}
    \tablecaption{Dust model parameters.\label{tab:modpara}}
    \tablewidth{0pt}
    \tablehead{
        \colhead{$f_{\rm vac}$} & \colhead{$f_{\rm carb}\left/f_{\rm sil}\right.$} & \colhead{$\amin$} & \colhead{$\amax$} & \colhead{$\alpha$} \\
        \nocolhead{} & \nocolhead{} & \colhead{($\um$)} & \colhead{($\um$)} & \nocolhead{}
    }
    \decimals
    \startdata
        &   & 0.1 &      &        \\
    0.5 & 3 &   1 & 1000 & 2.5-4.1\\
        &   &  10 &      &        \\
    \hline
    \multirow{4}{*}{0.9} & \multirow{4}{*}{3} & 0.1 & \multirow{4}{*}{1000} & \multirow{4}{*}{2.5-4.1}\\
                         &                    &   1 &                       &                         \\
                         &                    &  10 &                       &                         \\
                         &                    & 100 &                       &                         
    \enddata
\end{deluxetable}

\begin{figure*}[ht!]
    \plotone{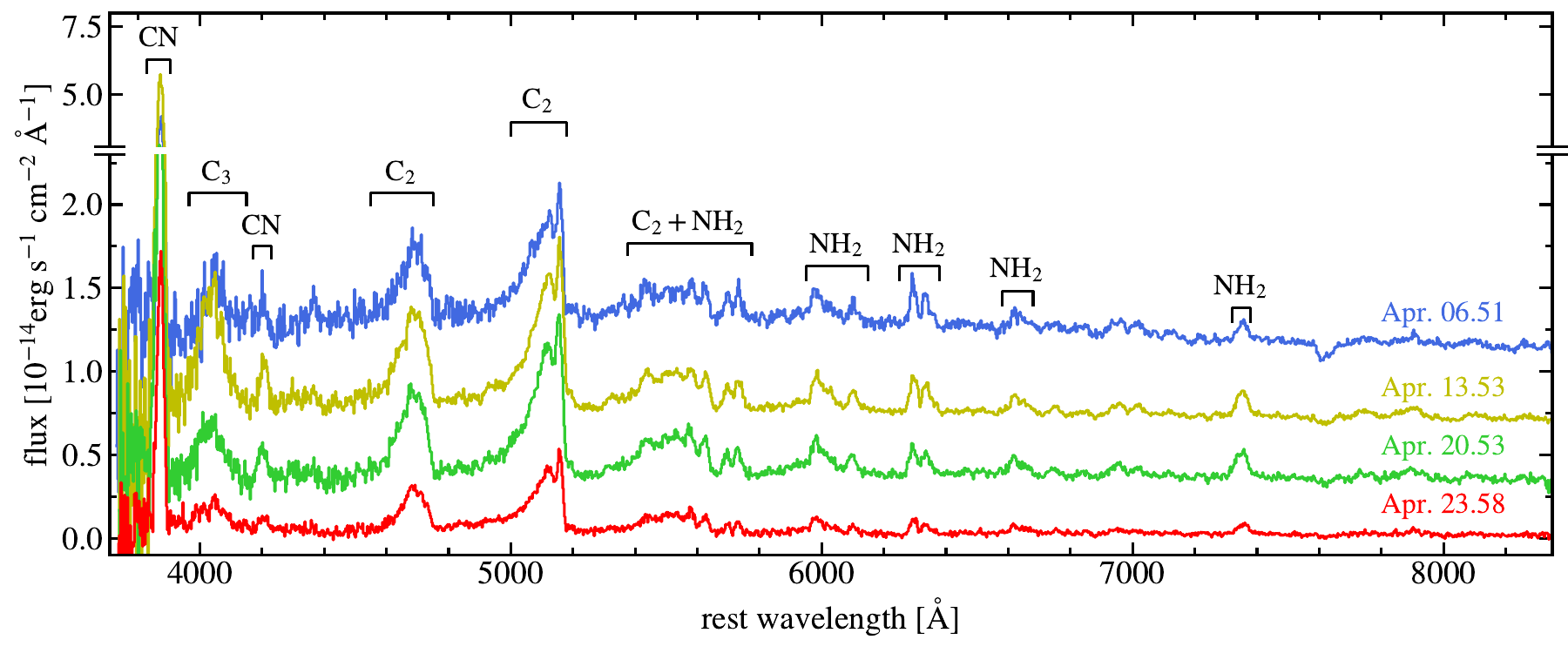}
    \caption{\footnotesize
    Narrow-aperture BFOSC spectra of C/2019 Y4 (ATLAS) obtained on 2020 April 6.51 (blue line), April 13.53 (yellow line), April 20.53 (green line), and April 23.58 (UT; red line). For comparison, the spectra are vertically shifted by some constants. Strong emission bands are marked. \label{fig:spec}}
\end{figure*}

\vspace{5cm}
\bibliography{19y4}{}
\bibliographystyle{aasjournal}



\end{document}